How and why communications industry suppliers get
"squeezed out" now, and the next phase

L-F Pau

# ERASMUS RESEARCH INSTITUTE OF MANAGEMENT

## REPORT SERIES
*RESEARCH IN MANAGEMENT*

| BIBLIOGRAPHIC DATA AND CLASSIFICATIONS | | |
|---|---|---|
| Abstract | The communications systems, terminals, software and deployment service, industries, have undergone the past ten years a significant technological internal evolution and external revolution at customer end (such as Internet, Mobile networks and terminals, Broadband,..). Very little management research has studied their financial survivability irrespective of changes in demand volumes in the present technological /organizational cycle. This paper analyzes the implications of genuine open mandated communications standards, of higher product volumes, of very high R&D, of the larger use of sourced /purchased technologies, and of contract manufacturing. The methodology used is equilibrium analyses. Two specific areas will be mentioned as examples. The paper also shows how eventually those industries in a later cycle will bounce back. | |
| Library of Congress Classification (LCC) | 5001-6182 | Business |
| | HE 9713+ | Wireless Telephone |
| Journal of Economic Literature (JEL) | M | Business Administration and Business Economics |
| | L 96 | Telecommunication |
| European Business Schools Library Group (EBSLG) | 85 A | Business General |
| | 55 D | Communication Techniques |
| Gemeenschappelijke Onderwerpsontsluiting (GOO) | | |
| Classification GOO | 85.00 | Bedrijfskunde, Organisatiekunde: algemeen |
| | 05.42 | Telecommunicatie |
| Keywords GOO | Bedrijfskunde / Bedrijfseconomie | |
| | Draadloze Communicatie, Telecommunicatie-industrie, Bedrijfsprocessen, Intellectueel eigendom | |
| Free keywords | Communication's industry, Communicatons industry suppliers, Business processes, Intellectual property, Competence, Customer bases | |

# How and why communications industry suppliers get "squeezed out" now, and the next phase .

L-F Pau, Draft

**Abstract** : The communications systems, terminals , software and deployment service, industries ,have undergone the past ten years a significant technological internal evolution and external revolution at customer end (such as Internet ,Mobile networks and terminals, Broadband,..). Very little management research has studied their financial survivability irrespective of changes in demand volumes in the present technological /organizational cycle .  This paper analyzes the implications of genuine  open mandated communications standards ,of higher product volumes , of very high R&D , of the  larger use of sourced /purchased technologies ,and of contract manufacturing . The methodology used is  equilibrium analyses . Two specific areas will be mentioned as examples .The paper also shows how eventually those industries in a later cycle will bounce back .

**Keywords :** Communication's industry , Communicatons industry suppliers , Business processes , Intellectual property , Competence , Customer bases

## Introduction

Network infrastructure as well as appliance/terminals suppliers , but also communications software , semiconductor , optical components producers ,and installation/deployment/management / billing service companies have over the past 10-15 years undergone simultaneously four radical external revolutions :
-the emergence and progressive domination of new technologies , such as Internet , wireless , broadband , and new semiconductor processes
-the deregulation and fragmentation of their customer base (public operators , enterprise networks
-the emergence of developing countries as major markets with leading edge demands
-changes in financing principles for suppliers (using "vendor financing") as well as their customers ( venture-like financing , heavy debt financing)

Whereas the above is widely researched , far less attention has gone into the operations and internal processes at these suppliers . At best , there has been a scattered debate as to what the "core business" and the "core competencies " were , with no clear conclusions emerging .

# 1.The intrinsic capital and knowledge bases of Communications industry suppliers

Historically , the base for most communications industry suppliers were end-to-end supply to monopolistic customers of telephone systems (from telephone access, to transmission , switches and network management) ,and progressively also of X25 packet communication systems (data terminals , routers ,and network management) .
While the complexity was significant , often because of many national/localized adaptations (signaling , electrical characteristics ,internal organization at the client's..) ,orders were still for networks or their capacity expansion or coverage .In almost all cases, manufacturing was an integral and profitable part of the supplier's organization .

With the deregulation of telephone sets and of data terminals , emerged the first time the question whether the systems' suppliers should manufacture and design these access terminals .In most cases , the decisions were to spin-off or outsource such manufacturing and very quickly also their design .

Ever since, the trend has accelerated to affect the:
- outsourcing of almost all manufacturing of system's nodes ,
- the co-design of key semiconductor or optical devices , shared between the communications industry and the semiconductor industry or the design houses
- the commercial acquisition and customization of large parts of the operating systems and middleware
- sometimes immoderate uses of consultants not just as buffer work capacity , but for key specification , design ,testing or other tasks

This leads to identify three essential capital bases for communications industry suppliers, in addition of course to the financial capital defined in company accounts :

a)The Intellectual capital and know-how capital , represented legally by inventions, patents , trademarks , process know-how and similar , and for which legal filing, protection and review mechanisms exists backed up by trade agreements and treaties

b)The competence capital ,which is the subjective set of skills, experience and contacts which staff brings to a company , and which the learning-by-doing process enhance as well as some occasional company initiated competence development activities .In this competence capital should be featured key staff ,and experts , who have a combination of higher creative skills, implementation skills and communication skills to their co-workers ,and sometimes to the community at large .Competence capital evaluation is a complex process, but it must obey to a cumulative process over time , with leakage linked to staff departure , to competence decay ,and sometimes to external factors who may make worthless some specific skills .

c)The <u>customer trust capital base</u> , which is not to be confused with the bilateral trade between a supplier and one of its customers , but with the depth , duration , credibility and trust of such a customer towards its suppliers . In one word it could be called the "depth of the relationship" .One trend has increased the customer trust base, which is the erosion of research and most technical activities at communications operators , who then have to rely more and more on their suppliers for their technical and some
other strategic choices .A reverse trend has been that, with the emergence of open standards in middleware ,interconnect and other API's , a communications operator or enterprise networking supplier can now in principle switch suppliers provided the new suppliers meet the same interface standards as the incumbent suppliers . The "depth of the relationship" is also always the compounded result of the relative internal forces, both at operator/enterprise end, as well as supplier end , between technical, operations, purchasing and financial decision makers . One clear trend however has been that, while this customer trust base is also a cumulative process, its rate of change (up or down) has vastly accelerated recently.

It should be noted that , for communications industry suppliers, the financial capital structure alone fails to give concrete indications as to the real value and survivability of  such suppliers . At best the financial capital structure and the cash flow allow to gage the pricing and cost base and the management of receivables ,as well indirectly as product/project/resource  management quality  .

## 2.The predators

As the business records of the past 10-15 years have shown ,and apart from the special case of mergers and acquisitions of and amongst communications industry start-ups , the consolidation has not been so much within communications industry suppliers .  One main reason has been heterogeneous elements  amongst the competence capital , the customer trust base and sometimes geographical localization and cost structures of the same .

Therefore predators have been from other sectors and the major strategic motivations have been :

a)for component suppliers (hardware, semiconductor , software , packaging, ...) to move up the value chain to grasp sub-systems or systems in growing communication offerings

b) for IT or computer industry suppliers to capitalize on high , but falling , computer sub-systems volumes ,to enhance them for the communications service needs ,and, in that process, to improve the quality and dependability of computer equipments . In some cases, such IT or computer industry suppliers attempt to achieve a monopoly in selected network architecture nodes (application servers , authentification servers , base station controllers , ....)

c) for consultants to occupy the place left open by the communications operators /enterprises in the field of service creation and service management ,as the diversity of these has been growing very fast with new technologies and open standards

d) for contract manufacturers to absorb the plants and production staff of communications industry suppliers ,although they had not realized that they would have a hard time going beyond bill-of-materials skills to reach architectural know-how and IPR without long lasting investments

e)for software products industry to enlarge its product portofolio to include major high-value packages sold by the communications industry suppliers (such as network management, billing systems, customer-care systems , messaging, security software  ) or used by these for their internal  processes (ERP, documentation , CAD , test systems)

f) for all the above to make the communications industry supplier bear the cost of risk research , of pre-standardization work, of standardization/lobbying ,of integration into very complex networks and systems , and of high cost initial engineering , productization and debugging

For these predators , two mechanisms have or are being used :

-merger and acquisitions , typically of communications industry supplier's plants , but also of non-core product lines or divisions ; in most cases , these deals have been carried out with financial valuation methods (or palettes thereof) which by and large have neglected totally the intellectual capital , the competence capital ,and the customer trust capital

-a "strangling approach " using technical dependencies as trigger events ,and business models thereafter ,and finally "strategic agreements" aiming at competence/IPR/customer trust  transfer from the communications industry supplier to a predator ;the core of this approach is precisely to take over the intellectual capital, the competence capital and/or the trust capital

## 3. Case (A) : A software industry predator

 In this Case, as well as in the following, we will , for the "strangling process" itemize the process and its impact on the intellectual , competence and customer trust capitals of the communications industry supplier.

The Case pertains to a software industry player, supplying a middleware component to a communications industry supplier .The general strategy used by the software supplier is a one of reducing the credibility of the communications suppliers, and to deploy a business model whereby communications systems suppliers and users are both being "taxed" , although alternative technologies exist ,some standardized by open ISO processes . The approach used by this software industry predator is also dual :

-merger and acquisition of start-up's and some other communications industry operations , including hardware ; it is interesting to note that most of these activities have failed , less because of external demand weaknesses, than because of internal fights where communications enhanced technologies and products were "forced" to report to established product lines , leading to bad technical solutions , de-motivated new employees ,and total lack of trust with the customers of the acquired party

-"strangling" of the communications industry supplier by the following process :

Table 1: Software industry predator process

| STEP | PROCESS | IMPACT on Intellectual capital | IMPACT on Competence capital | IMPACT on Customer trust capital |
|---|---|---|---|---|
| 1 | Specify alone ,and brand a software of rather simple design , but with novel functionalities | | Competition on innovation is fair | |
| 2 | Apply substantial marketing power to publicize this brand ,not the least with communications operators , so the communications industry suppliers would be forced to meet their customer's "demands" or inquiries | | | Negative as it reduces sometimes credibility of communications supplier, especially if they have researched the same functionality |
| 3 | Apply substantial lobbying at technical management level at communications industry supplier's , essentially to tell these management levels that they would be "left behind" if they don't commit to this software | | Negative as it corresponds to criticize internal resources developing same | |
| | | | functionality | |
| 4 | Charge a developer's license ,and impose a trademark license for the brand | Negative if license prevents from doing internal work in same area | | Negative as operator is used to only the communications supplier's brand |
| 5 | Impose the software vendor , or a proxy to it , as the sole party who can develop conformance test suites in view of use of the brand by communications supplier in his products when they embed this software | | | Negative as operator is used to open public conformance suites |
| 6 | Charge a license fee for test suite(s),and force resource allocation to the task | | Negative as resource could be used better to enhance IPR capital | |
| 7 | Charge a royalty on software when embedded into supplier's products | | | Positive if royalty cost is less than depreciation of a communications supplier's alternative solution |
| 8 | Charge a test suite license (in view of the branding) and a distribution license to operator on products received from supplier | | | Negative as this time operator will ask communications supplier why he forced this side-cost |
| 9 | Offer a R&D collaboration to supplier, with free evaluation license by software company on communications supplier's add-on's | Negative as no pay-back on IPR | Negative as freedom is limited | |

| 10 | Include same functionality as obtained under Step 9, but under independent implementation into software product | Negative as functionality is hard to protect legally | Positive for those who see their ideas diffuse | |
| 11 | Charge a license to communications supplier for enhancement 10. Which originates in 9.,and under software company's brand | | Negative as communications supplier have to pay twice | Negative |
| 12 | Integrate communications software protocols into software product (signalling , protocols , OSS ,..) | Negative as communications suppliers must protect all implementations of communications stacks | Negative as staff is left only to maintain "old" versions " of communications software | Negative as "image" in communications software field shifts to software industry |

## 4.Case (B): a contract manufacturer predator

In this Case, as well as in the following, we will , for the "strangling process" itemize the process and its impact on the intellectual , competence and customer trust capitals of the communications industry supplier.

The Case pertains to a contract manufacturer, with its roots in mechanical assembly , who carried out an aggressive set of opportunities to take over (in general for free) some manufacturing operations of a diversity of weakened communications industry suppliers . The goal was to migrate from contract manufacturing to design and test services , to acquire design know-how,and to produce unbranded communications products competing with those of its original customer .

Thus here again ,the predator has a dual approach :

-merger and acquisitions : essentially , the predator off-loaded the communications industry supplier of social costs and underutilized production capacities, got some initial orders as part of the deal, but never committed to manufacturing or engineering price levels ;the result is that , after some time, the same production cost for the same item goes substantially up with no quality guarantees possible

-"strangling" process initiated under the excuse of shared traditions, design methodologies and better purchasing power at component level ; the target was to carry out test, low volume /prototyping assembly , engineering services , as well as to help the communications industry supplier with faster industrialization services for its end customers .Thus the process and its impacts for the communications industry supplier were the following :

Table 2: Contract manufacturer industry predator process

| STEP | PROCESS | IMPACT on Intellectual capital | IMPACT on Competence capital | IMPACT on Customer trust capital |
|---|---|---|---|---|
| 1 | Offer low volume manufacturing (prototypes ,...) | | Faster and fabless delivery of prototypes | Faster access to prototypes for evaluation |
| 2 | Offer test design and test execution services , often with tools and fixtures inherited from communications industry supplier | | Negative as concurrent design and testing are key for enhancing designs | |
| 3 | Offer integration services | Negative as architectural specifications , tricks and know how got transferred in effect without real IPR protection | Negative as integration staff leaves and joins contract manufacturer | Negative as communications industry supplier looks as incapable of being a systems integrator |
| 4 | Offer volume component purchasing services | Negative as component supplier data and choices get compromised | Negative on sourcing staff | |

| 5 | Renegotiate contract manufacturing and service contracts to add contract manufacturer's so-called "own" IPR and licensing terms (know how not backed up by own investments but happening by Step 3) | Negative | | Negative as operator buys less from communications supplier, because of less trust in contract manufacturer's abilities and concern for operator |
|---|---|---|---|---|
| 6 | Offer to third parties unbranded down scale products competing with original communications supplier's | | Negative as staff's pride is diminished and some leave for contract manufacturer | Negative as operator may switch to unbranded cheaper product assuming its design to be dependable |
| 7 | Sell product line as part of a focussing strategy, often to lower cost manufacturers | Negative on royalties | Negative on staff which is often being made redundant both at communications supplier as well as contract manufacturer | Negative in general |

## 5. Case (C) : a semiconductor supplier predator

In this Case we will again, for the "strangling process" itemize the process and its impact on the intellectual, competence and customer trust capitals of the communications industry supplier.

In this Case, the strategy of the semiconductor supplier is to go up the value chain by claiming that its process and volume advantages will be of benefit to the communications industry supplier as well as to the operators, although he may fail to say that the architectures and price levels may be quite different and reduce the learning curve benefits .

The merger and acquisition process does not affect the communications industry supplier ,as the focus of the semiconductor supplier is on investing in start-up's and individuals who may allow it to leapfrog the communications industry's supplier's know how and competence in a generic area (a "platform")

The "strangling" process thereafter as the following steps :

Table 3: Semiconductor industry predator process

| STEP | PROCESS | IMPACT on Intellectual capital | IMPACT on Competence capital | IMPACT on Customer trust capital |
|---|---|---|---|---|
| 1 | Invest in start-up's and individuals with generic technologies at a low level, which are being forced via financial controls to abide with the semiconductor's processes, IPR and proprietary architectures | Lost opportunities to acquire cheap some potential relevant IPR | | |
| 2 | Organize a fully controlled industry forum with strict NDA rules ,and mandated dependency on the semiconductor supplier's prior architectures and proprietary standards | | Positive as staff asked to join individually may feel that they are important | |
| 3 | Carry out a heavy marketing campaign around yet-to-be products claiming industry backing by the other parties in the controlled industry forum ; lobby operators as to the future capabilities and cost benefits to them | Negative as architectural specifications , tricks and know how got transferred in effect without real IPR protection | Negative as staff is confused about either doing their own thing, or buying in | Negative as operators are confused as to why communications industry supplier charges more than they are told they should by semiconduc |

| | | | | tor vendor |
|---|---|---|---|---|
| 4 | Lobby technical and sourcing management of communications industry supplier as to cheaper prices ,volume advantages , for yet-to-be-seen and yet-to-be-priced products ; pressure from operator feedbacks ; holding back access to key components | Negative as no open standardization is possible and IPR is made dependent on semiconductor supplier's IPR | Negative on staff who sees R&D budgets and products curtailed to the benefit of external paper products | Positive with buyers ; negative with operator's technical resources who see delays ahead |
| 5 | Offer bundled components and sub-systems ,integrated by semiconductor vendor's systems divisions | Negative as architectural innovation vanes | Negative as resources are laid off because of operating costs , rarely objectively compared with the semiconductor" sub-system life-cycle costs to the communications supplier | Positive as communications supplier is seen as embarking on "industry standards" ,until life-cycle added costs become apparent to operator |
| 6 | Stop supplying components to communications systems supplier, who is forced to adopt sub-systems bundles ,and in effect withdraw from platforms area | No IPR generated around platform | Staff laid off | Negative as operator supports all trouble with new field support , higher prices ,etc |

## 6. Discussion of cases

It should first be noted that, for each of the three Cases (A-C) , there are more than one ,often minimum two suppliers applying such processes towards communications industry suppliers . It is not feasible to disclose here these predator suppliers .

Next , the quantifiable impact on the IPR capital , competence capital ,and customer trust capital, of these processes depend very much on the patent portofolio mix, of the product/system/service mix , on the personnel age /experience / skills distributions ,and finally of the technology bases and established customer bases .It is also difficult to quantify this impact in absolute terms as the three types of capital do not have established measurements recognized by accounting standards

Nevertheless, the three Cases are representative of needs of most communications industry suppliers , so that they are affected by these three processes altogether due to the global character of the operations of most predators and of the communications suppliers themselves .

An attempt is therefore made to evaluate the relative impact of the three processes, using publicly available elements pertaining to the patent portofolio , personnel resource distribution ,and the announced product plans and contracts over the period mid 2002-end 2004 ,by relatively large communications industry suppliers , which :
-either have a fairly diversified communications product/service portofolio : Lucent , Nortel , Siemens (ICN+ICM) , Ericsson , Alcatel (excluding space activities and others) ,NEC (Communications Div)
-or have a more focussed communications product/service portofolio : Nokia , Marconi , Tellabs

The result of the evaluation are relative impact ranges , over the mid 2002-end 2004 period, with the mid-2002 period as a base for the three forms of capital :

Table 4 : Cumulative relative impact table from predatory processes aiming at communications industry suppliers , in % , mid-2002 to end 2004 , with base mid 2002

| PROCESS | Relative impact on IPR capital | Relative impact on competence capital | Relative impact on client trust capital |
| --- | --- | --- | --- |
| Case A: software | 2-5 % | 0- 7 % | 0-4 % |
| Case B : contract manufacturing | 4-16 % | 2-18 % | 6-14 % |
| Case C: semiconductor | 8-25 % | 5-20 % | 10-25 % |
| Minimum (maximum) cumulated relative impact in % , 2002-2004 | 14 % (46 %) | 7 % (45 %) | 16 % (43 %) |

Obviously the maximum relative cumulated impacts are unrealistic, but the very fact that they are all three in the 45 % domain is perceptually indicative of both a tolerance level ,or of an awakening level , at the communications industry supplier's . The debate as to the ratio of dependency towards non-communications industry suppliers for core technologies , skills and customer relations ,is very rarely debated , except for:
- purchasing/sourcing departments who saw in positive eyes higher dependencies (up to 50 %) if the short term costs offer short term financial benefits on the operating costs
- financial management who see that the non-communications industry suppliers, with such predatory processes, reap the higher margins elements and leave the resource costly and low margin activities to the communications industry suppliers
- technical managers who would normally fight against any "N.I.H" level higher than the very minimum, in the 10-15 % range ;if they seem to be alerted to the situation it is interestingly enough because the estimated Minimum cumulative relative impact levels from the Table , have reached the alarm level for this category of decision makers

## 7. The communications industry suppliers' defense and offensive strategies

Here are short formulations of such strategies, the relative impacts of which can also be estimated via an impact table like Table 4 .

a) Mandate open public standards (ISO process) compliance in all sourced products or technologies :

   The growing importance of 3GPP , IETF , OMG , OMA is an illustration of this , as it allows in theory to switch component vendors , but not systems vendors or know how suppliers (such as consultants)

b) Cooperative funding , or "pool companies" to develop generic components to communications industry specifications , and with test and integration by the communications industry

   Symbian , power module industry ,and in effect semiconductor foundries , are examples thereof

c) Collaboration of communications industry suppliers in joint specifications beyond standards (or prior to standards are published)

   This is a difficult process due to anti-trust policies , but recent political initiatives have pointed at the benefits of such approaches for the industry at least on regional basis (European Union, China/Taiwan )

d) Involving operators and other communications industry clients in management and technology roadmap alignment processes

This process is extended to address not just service level features but also architectural elements including in effect choices of supplier alternatives .

e) Timely and effective patent and IPR swap agreements with royalty payments or compensation payments

   This mechanism has long been used for essential patents used mandated for architectural compliance in standards, but being extended to outright IPR licensing businesses to selected other communications industry suppliers (e.g. agreements around Bluetooth technology , 3G technology , handset platform technologies and industrialization packages )

f) Branding policy by communications industry suppliers (individually or collectively) of communications software applications or packages ,instead of letting the software , IT or semiconductor industry do it

   This mechanism both keeps the customer trust image ,but is also a way to multiply the adoption of IPR or API's owned by the communications industry . Examples include System availability Forum , BREW ( by Qualcomm) , etc..

g) Aggressive personnel policies offering technical career tracks and prestige

   This process ,probably initially tested at IBM and Bell Labs , has spread widely to the communications industry suppliers .However , are too often rewarded in this way old skills ,rather than novel areas or skills stimulating disruptive innovation and competivity .

h) Innovative business process ,tariff/charging/rating/ payment schemes benefiting the operators and supported by innovative business processes embedded inside the technical systems and products

   The 3G Mobile Internet,and IPv6 areas start to offer such cases , although the staff involved at the communications industry supplier's is rarely looked positively upon for such non-technical innovation . Often ,this work is left to consultants and the competence capital and customer trust capital decrease

## CONCLUSION

The communications industry supplier's face far deeper challenges than the financial situation and demand evolution . The consolidation processes in this industry cannot ignore the predatory ambitions by other industries , and the identification of mismatch between the suppliers themselves and with others in terms of IPR , competence and customer trust , are necessary to enable success over time and continued innovation in the best interest of end users .

# Publications in the Report Series Research* in Management

ERIM Research Program: "Business Processes, Logistics and Information Systems"

2003

*Project Selection Directed By Intellectual Capital Scorecards*
Hennie Daniels and Bram de Jonge
ERS-2003-001-LIS
http://hdl.handle.net/1765/265

*Combining expert knowledge and databases for risk management*
Hennie Daniels and Han van Dissel
ERS-2003-002-LIS
http://hdl.handle.net/1765/266

*Recursive Approximation of the High Dimensional max Function*
Ş. İl. Birbil, S.-C. Fang, J.B.G. Frenk and S. Zhang
ERS-2003-003-LIS
http://hdl.handle.net/1765/267

*Auctioning Bulk Mobile Messages*
S.Meij, L-F.Pau, E.van Heck
ERS-2003-006-LIS
http://hdl.handle.net/1765/274

*Induction of Ordinal Decision Trees: An MCDA Approach*
Jan C. Bioch, Viara Popova
ERS-2003-008-LIS
http://hdl.handle.net/1765/271

*A New Dantzig-Wolfe Reformulation And Branch-And-Price Algorithm For The Capacitated Lot Sizing Problem With Set Up Times*
Zeger Degraeve, Raf Jans
ERS-2003-010-LIS
http://hdl.handle.net/1765/275

*Reverse Logistics – a review of case studies*
Marisa P. de Brito, Rommert Dekker, Simme D.P. Flapper
ERS-2003-012-LIS
http://hdl.handle.net/1765/277

*Product Return Handling: decision-making and quantitative support*
Marisa P. de Brito, M. (René) B. M. de Koster
ERS-2003-013-LIS
http://hdl.handle.net/1765/278

---

\*  A complete overview of the ERIM Report Series Research in Management:
http://www.erim.eur.nl

ERIM Research Programs:
LIS   Business Processes, Logistics and Information Systems
ORG  Organizing for Performance
MKT  Marketing
F&A  Finance and Accounting
STR  Strategy and Entrepreneurship